\documentstyle[11pt]{article}
\begin{document}

\title{{\bf Time reparameterization in Bianchi type I spinor cosmology}}
\author{B. Vakili\thanks{%
email: b-vakili@sbu.ac.ir} and H. R. Sepangi\thanks{%
email: hr-sepangi@sbu.ac.ir} \\
{\small Department of Physics, Shahid Beheshti University, Evin, Tehran
19839, Iran}}
\maketitle

\begin{abstract}
The problem of time reparameterization is addressed at both the
classical and quantum levels in a Bianchi-I universe in which the
matter source is a massive Dirac spinor field. We take the scale
factors of the metric as the intrinsic time and their conjugate
momenta as the extrinsic time. A scalar character of the spinor
field is identified as a representation of the extrinsic time. The
construction of the field equations and quantization of the model
is achieved by solving the Hamiltonian constraint after time
identification has been dealt with. This procedure leads to a true
Hamiltonian whose exact solutions for the above choices of time
are presented.\vspace{5mm}\newline PACS numbers: 04.20.-q,
04.50.+h, 04.60.-m
\end{abstract}

\section{Introduction}

Standard cosmological models based on classical general relativity
have no convincing precise answer for the presence of the
so-called ``Big-Bang'' singularity. Any hope of dealing with such
singularities would be in vein unless a reliable quantum theory of
gravity can be constructed. In the absence of a full theory of
quantum gravity, it would be useful to describe the quantum states
of the universe within the context of quantum cosmology,
introduced in the works of DeWitt \cite{1} and later Misner
\cite{2}. In this formalism which is based on the canonical
quantization procedure, one first freezes a large number of
degrees of freedom and then quantizes the remaining ones. The
quantum state of the universe is then described by a wave function
in the mini-superspace, a function of the 3-geometry of the model
and matter fields presented in the theory, satisfying the
Wheeler-DeWitt (WD) equation. In more recent times such works have
been the focus of an active area of research with different
approaches, \cite{3}-\cite{11}, see also \cite{12} for a review.
In references \cite{13}, canonical quantization is applied to many
models with different matter fields as the sources of gravity.

As are well known, quantum cosmology suffers from a number of
problems, namely the construction of the Hilbert space to define a
positive definite inner product of the solutions of the WD
equation, the operator ordering problem and also most importantly,
the problem of time. The wave function in the WD equation is
independent of time, {\it i.e.} the universe has a static picture
in this scenario. This problem was first addressed in \cite{1} by
DeWitt himself. However, he argued that the problem of time should
not be considered as a hinderance in the sense that the theory
itself must include a suitable well-defined time in terms of its
geometry or matter fields. In this scheme time is identified with
one of the characters of the geometry, usually the scale factors
of the geometry and is referred to as the intrinsic time, or with
the momenta conjugate to the scale factors, or even with a scalar
character of matter fields coupled to gravity in any specific
model, known as the extrinsic time.

In general, the crucial problem in canonical quantum gravity is
the presence of constraints in the gravitational field equations.
Identification of time with one of the dynamical variables depends
on the method we use to deal with theses constraints. Different
approaches arising from these methods have been investigated in
detail in \cite{14}. The issue of time in canonical general
relativity is also extensively discussed in \cite{e}. As has been
discussed in \cite{14}, time may be identified before or after
quantization has been done. There are approaches, on the other
hand, in which time has no fundamental role. For a more modern
review of the problem of time and other related problems in
quantum cosmology see \cite{15}. The details of time
identification procedure in terms of various dynamical variables
of the theory before quantization is done has been investigated in
\cite{16} where a Robertson-Walker universe filled with a scalar
field is quantized. Also in \cite{17} a choice of time in terms of
a massless scalar field is discussed in a Bianchi-I classical
cosmology based on the method developed in \cite{18}.

One of the common candidates of time in the above works is the
matter field present in the theory, that is, time is identified
with a scalar character of matter. In the case of a scalar field
as the source of gravity, the scalar field itself can play the
role of time as is the case in \cite{16} and \cite{17}. Another
matter field which has occasionally been studied in the
literatures is the massless or massive spinor field as the source
of gravity. In general, theories studying spinor fields coupled to
gravity result in Einstein-Dirac systems which are not easy to
solve. The quantized Robertson-Walker or Bianchi-I universe filled
with a spinor field are studied in \cite{19}-\cite{21}. For a
general discussion on the possibility that classical homogeneous
spinor fields might play the role of matter in cosmology, the
reader is referred to \cite{22}.

In this paper we deal with classical and quantum cosmology of a
model in which a classical massive spinor field is coupled to
gravity in a Bianchi type I space-time. What we mean by a
classical spinor field is a set of four complex-valued space time
functions that transform according to the spinor representation of
the Lorentz group. The existence of such fields is crucial in our
work since in spite of fact that fermions are described by
quantized spinor fields which do not have a classical limit, we
assume such classical fields exist and use them as matter fields
coupled to gravity. A possible justification for the existence of
classical spinor fields is given in the appendix of reference
\cite{22}. To identify time, we have adopted the same procedure as
in \cite{16}, that is, after choosing a time parameter, we solve
the Hamiltonian constraint equation to obtain a minimal true
Hamiltonian. We then use this reduced Hamiltonian to construct the
classical field equations. Also, to quantize the model we  use the
operator form of the resulting Hamiltonian in the Schr\"{o}dinger
picture. The reason that we have used the Schr\"{o}dinger picture
for quantizing the system and not the WD picture is that time
naturally enters the Schr\"{o}dinger equation. As is well known in
the WD formalism, in quantizing a cosmological system, there is no
a priori definition of time. This is quite natural and stems from
the nature of the WD equation and is a reflection of the gauge
invariance with respect to the choice of coordinates in the
classical theory. The intrinsic times which we have chosen in this
paper are related to the scale factors of the metric, with their
conjugate momenta chosen as the extrinsic time. We have shown that
there is also an identification of the extrinsic time in terms of
the scalars constructed with spinor fields. The exact solutions of
the classical and quantum cosmology corresponding to each choice
of time are presented. The complicated form of some of these
solutions show that the evolution of the universe with respect to
different intrinsic or extrinsic times is a non-trivial
undertaking.

\section{The classical model}

We start with a space-time metric of the form \footnote{We work in
the units where $c=\hbar=16\pi G=1$}
\begin{equation}  \label{A}
ds^2=-N^2(t)dt^2+a^2(t)dx^2+b^2(t)dy^2+c^2(t)dz^2,
\end{equation}
which describe a Bianchi type I universe with scale factors
$a(t)$, $b(t)$ and $c(t)$ in the $x$, $y$ and $z$ directions
respectively, with $N(t)$ being the lapse function. This metric is
the simplest anisotropic and homogeneous cosmological model which,
upon making the scale factors equal, becomes the flat
Robertson-Walker metric. Such space-times have an Abelian symmetry
group of
translations with Killing vector fields ${\bf \xi}=(\partial_x,\partial_y,%
\partial_z)$. Of course, all the structure constants of such a symmetry
group are zero. The properties of such a space time is studied in
many works, see for example \cite{b} and \cite{d} and the
references therein. The scalar curvature corresponding to metric
(\ref{A}) is
\begin{equation}  \label{B}
{\cal R}=\frac{2}{N^2}\left(\frac{\ddot{a}}{a}+\frac{\ddot{b}}{b}+\frac{%
\ddot{c}}{c}+ \frac{\dot{a}\dot{b}}{ab}+\frac{\dot{b}\dot{c}}{bc}+\frac{\dot{%
c}\dot{a}}{ca} -\frac{\dot{N}\dot{a}}{Na}-\frac{\dot{N}\dot{b}}{Nb}-\frac{%
\dot{N}\dot{c}}{Nc}\right),
\end{equation}
where a dot represents differentiation with respect to $t$. To
construct the field equations, let us start with the action
\begin{equation}  \label{C}
{\cal S}=\int (L_{grav}+L_{matt})\sqrt{-g}d^4x,
\end{equation}
where
\begin{equation}  \label{D}
L_{grav}={\cal R}-2\Lambda,
\end{equation}%
is the Einstein-Hilbert Lagrangian for the gravitational field
with cosmological constant $\Lambda$, and $L_{matt}$ represents
the Lagrangian of the matter source which we assume to be a
classical massive free spinor field. As we mentioned in
introduction, for the bulk of this paper we shall consider the
spinor fields as classical objects, {\it i. e.}  four
complex-valued space time functions and not a set of Grassmanian
variables. In fact spinors in quantum field theory are four
operator-valued fields which act on the corresponding Hilbert
space and obey the Dirac equation (in flat space time)
\begin{equation}\label{re1}
(\gamma^{\mu}\partial_{\mu}-m)\psi=0.
\end{equation}
However, in what follows, following \cite{22}, we shall interpret
a classical spinor field, that is, a set of four complex-valued
functions as
\begin{equation}\label{re2}\psi_{cl}=\langle s\mid\psi\mid s\rangle,
\end{equation}
where $\mid s\rangle$ is an appropriate physical state. Now, we
can see that the expectation value of a spinor in a physical state
is a complex number and not a Grassmanian number. Also taking the
expectation value of the Dirac equation (\ref{re1}) yields
\begin{equation}\label{re3}
(\gamma^{\mu}\partial_{\mu}-m)\psi_{cl}=0,
\end{equation}
which means that the classical spinor fields also obey the Dirac
equation. In what follows, by $\psi$ we mean the classical field
$\psi_{cl}$ and omit the subscript $cl$ from now on. For a more
extensive discussion of the properties of such classical fields
see \cite{22}.

As usual, the Dirac equation describing the dynamics of a spinor
field $\psi$ can be obtained from the Lagrangian
\begin{equation}  \label{E}
L_{matt}=\frac{1}{2}\left[\bar{\psi}\gamma^{\mu}(\partial_{\mu}+
\Gamma_{\mu})\psi-\bar{\psi}(\overleftarrow{\partial_{\mu}}
-\Gamma_{\mu})\gamma^\mu\psi \right]-V(\bar{\psi},\psi),
\end{equation}
where $\gamma^{\mu}$ are the Dirac matrices associated with the space-time
metric satisfying the Clifford algebra $\{\gamma^{\mu},\gamma^{\nu}\}=2g^{%
\mu \nu }$, $\Gamma_{\mu}$ are spin connections and $V(\bar{\psi},\psi)$ is
a potential describing the interaction of the spinor field with itself. In
the case of a free spinor field of mass $m$ we have $V(\bar{\psi},\psi)=m%
\bar{\psi}\psi$. The $\gamma^{\mu}$ matrices are related to the flat Dirac
matrices, $\gamma^a$, through the tetrads $e^a_{\mu}$ as follows
\begin{equation}  \label{F}
\gamma^{\mu}=e^{\mu}_{a}\gamma^a, \hspace{.5cm} \gamma_{\mu}=e^a_{\mu}%
\gamma_a.
\end{equation}
For the metric (\ref{A}) the tetrads can be easily obtained from their
definition, that is $g_{\mu \nu}=e^a_{\mu}e^b_{\nu}\eta_{ab}$, leading to
\begin{equation}  \label{G}
e^a_{\mu}=\mbox{diag}(N,a,b,c),\hspace{.5cm} e^{\mu}_a=\mbox{diag}%
(1/N,1/a,1/b,1/c).
\end{equation}
Also, the spin connections satisfy the relation
\begin{equation}  \label{H}
\Gamma_{\mu}=\frac{1}{4}g_{\nu\lambda}(\partial_{\mu}e^{\lambda}_a+
\Gamma^{\lambda}_{\sigma\mu}e^{\sigma}_a)\gamma^{\nu}\gamma^a.
\end{equation}
Thus, for the line element (\ref{A}), use of (\ref{F}) and (\ref{H}) yields
\begin{equation}  \label{I}
\Gamma_0=0,\hspace{.5cm}\Gamma_1=-\frac{\dot{a}}{2N}\gamma^0 \gamma^1,%
\hspace{.5cm} \Gamma_2=-\frac{\dot{b}}{2N}\gamma^0
\gamma^2,\hspace{.5cm} \Gamma_3=-\frac{\dot{c}}{2N}\gamma^0
\gamma^3.
\end{equation}
Here $\gamma^0$ and $\gamma^i$ are the Dirac matrices in Minkowski
space and we have adopted the following representation \cite{23}
\begin{equation}  \label{J}
\gamma^0=\left(%
\begin{array}{cc}
-i & 0 \\
0 & i \\
\end{array}%
\right),\hspace{.5cm} \gamma^i=\left(%
\begin{array}{cc}
0 & \sigma^i \\
\sigma^i & 0 \\
\end{array}%
\right).
\end{equation}
The final remark about Lagrangian (\ref{E}) is that consistency of
Einstein field equations with a spinor field as the matter source
in the background metric (\ref{A}) requires the spinor field
$\psi$ to be dependent on $t$ only, that is $\psi=\psi(t)$
\cite{20}.

The preliminary set-up for writing the action is now complete. Substituting (%
\ref{B}), (\ref{D}) and (\ref{E}) into (\ref{C}) and integrating
over the spatial dimensions, we are led to an effective Lagrangian
in the mini-superspace
$\{N,a,b,c,\psi,\bar{\psi}\}$\footnote{Although, it is not a
priori evident that the substitution of the ansatz for the metric
and the matter fields into the action leads to the correct
equations of motion, here, as we can see from the equations, this
is the case. This procedure is not correct when class B Bianchi
models are involved.}
\begin{equation}  \label{K}
{\cal L}=\frac{1}{N}\left(\dot{a}\dot{b}c+a\dot{b}\dot{c}+\dot{a}b\dot{c}%
\right)+\Lambda Nabc+\frac{1}{2}Nabc\left[\frac{1}{N}(\bar{\psi}\gamma^0
\dot{\psi}-\dot{\bar{\psi}}\gamma^0 \psi)-2V(\bar{\psi},\psi)\right].
\end{equation}
Variation of Lagrangian (\ref{K}) with respect to $\bar{\psi}$, $\psi$, $%
a$, $b$, $c$ and $N$ yields the equations of motion of the spinor and the
gravitational fields as
\begin{equation}  \label{L}
\dot{\psi}+\frac{1}{2}\left(\frac{\dot{a}}{a}+\frac{\dot{b}}{b}+ \frac{\dot{c%
}}{c}\right)\psi+N\gamma^0\frac{\partial V}{\partial\bar{\psi}}=0,
\end{equation}
\begin{equation}  \label{M}
\dot{\bar{\psi}}+\frac{1}{2}\left(\frac{\dot{a}}{a}+\frac{\dot{b}}{b}+ \frac{\dot{c}}{c}%
\right)\bar{\psi}-N\frac{\partial V}{\partial \psi}\gamma^0=0,
\end{equation}

\begin{equation}  \label{N}
\frac{\ddot{b}}{b}+\frac{\ddot{c}}{c}+\frac{\dot{b}\dot{c}}{bc}- \frac{\dot{N%
}}{N}\left(\frac{\dot{b}}{b}+\frac{\dot{c}}{c}\right )- \Lambda N^2=\frac{1}{%
2}N^2\left[\bar{\psi}\frac{\partial V}{\partial \bar{\psi}}+\frac{\partial V%
}{\partial \psi}\psi \right]-V(\bar{\psi},\psi),
\end{equation}

\begin{equation}  \label{O}
\frac{\ddot{c}}{c}+\frac{\ddot{a}}{a}+\frac{\dot{a}\dot{c}}{ac}-\ \frac{\dot{%
N}}{N}\left(\frac{\dot{c}}{c}+\frac{\dot{a}}{a}\right) -\Lambda N^2=\frac{1}{%
2}N^2\left[\bar{\psi}\frac{\partial V}{\partial \bar{\psi}}+\frac{\partial V%
}{\partial \psi}\psi \right]-V(\bar{\psi},\psi),
\end{equation}

\begin{equation}  \label{P}
\frac{\ddot{a}}{a}+\frac{\ddot{b}}{b}+\frac{\dot{a}\dot{b}}{ab}- \frac{\dot{N%
}}{N}\left(\frac{\dot{a}}{a}+\frac{\dot{b}}{b}\right)-\Lambda N^2=\frac{1}{2}%
N^2\left[\bar{\psi}\frac{\partial V}{\partial \bar{\psi}}+\frac{\partial V}{%
\partial \psi}\psi \right]-V(\bar{\psi},\psi),
\end{equation}
\begin{equation}  \label{W}
\frac{\dot{a}\dot{b}}{ab}+\frac{\dot{a}\dot{c}}{ac}+\frac{\dot{b}\dot{c}}{bc}%
-\Lambda N^2=-N^2 V(\bar{\psi},\psi).
\end{equation}
As a double check, one may obtain the above field equations from the Dirac
and Einstein equations, given as
\begin{equation}  \label{Q}
\gamma^{\mu}(\partial_{\mu}+\Gamma_{\mu})\psi-\frac{\partial V}{\partial
\bar{\psi}}=0,
\end{equation}
\begin{equation}  \label{R}
\bar{\psi}(\overleftarrow{\partial_{\mu}}-\Gamma_{\mu})\gamma^{\mu}+\frac{%
\partial V}{\partial \psi}=0,
\end{equation}
\begin{equation}  \label{S}
R_{\mu \nu}-\frac{1}{2}{\cal R}g_{\mu \nu}+\Lambda
g_{\mu\nu}=T_{\mu\nu}.
\end{equation}
For a homogeneous spinor field $\psi=\psi(t)$, equations (\ref{Q}) and (\ref%
{R}) are equivalent to (\ref{L}) and (\ref{M}) respectively. On
the other hand, the non-vanishing components of the Einstein
tensor for metric (\ref{A}) are
\begin{equation}  \label{T}
G_{00}=\frac{\dot{a}\dot{b}}{ab}+\frac{\dot{a}\dot{c}}{ac}+\frac{\dot{b}\dot{%
c}}{bc},
\end{equation}
\begin{equation}  \label{U}
G_{11}=-\frac{a^2\ddot{b}}{bN^2}-\frac{a^2\ddot{c}}{cN^2}+ \frac{a^2\dot{b}%
\dot{N}}{bN^3}+\frac{a^2\dot{c}\dot{N}}{cN^3} -\frac{a^2\dot{b}\dot{c}}{bcN^2%
},
\end{equation}
with the cyclic permutations $a\rightarrow b \rightarrow c$ on
$G_{11}$ giving the expressions for $G_{22}$ and $G_{33}$. Also
the components of the energy-momentum tensor for the spinor field
as the matter source can be obtained from the standard definition
\[
T_{\mu\nu}=2\frac{\partial L_{matt}}{\partial g^{\mu\nu}}-g_{\mu\nu}L_{matt},
\]
yielding
\begin{equation}  \label{V}
T_{00}=-N^2 V(\bar{\psi},\psi),\hspace{.5cm}T_{11}=\frac{1}{2}a^2\left(\bar{%
\psi}\frac{\partial V}{\partial \bar{\psi}}+\frac{\partial V}{\partial \psi}%
\psi\right)+V(\bar{\psi},\psi),\hspace{.5cm}T_{ij}=T_{0i}=0.
\end{equation}
Again the above mentioned permutations on $T_{11}$ lead to $T_{22}
$ and $T_{33}$. Substitution of these results into Einstein equations (\ref%
{S}) yields the same equations as (\ref{N})-(\ref{W}).

Let us now construct the Hamiltonian for our model. The momenta
conjugate to the dynamical variables are
\begin{equation}  \label{X}
p_a=\frac{\partial {\cal L}}{\partial \dot{a}}=\frac{1}{N}(\dot{b}c+b\dot{c}%
),\hspace{.5cm} p_b=\frac{\partial {\cal L}}{\partial \dot{b}}=\frac{1}{N}(a%
\dot{c}+c\dot{a}),\hspace{.5cm} p_c=\frac{\partial {\cal L}}{\partial \dot{c}%
}=\frac{1}{N}(a\dot{b}+\dot{a}b),
\end{equation}
and
\begin{equation}  \label{Y}
p_{\psi}=\frac{\partial {\cal L}}{\partial \dot{\psi}}=\frac{1}{2}abc\bar{%
\psi}\gamma^0,\hspace{.5cm} p_{\bar{\psi}}=\frac{\partial {\cal L}}{\partial
\dot{\bar{\psi}}}=-\frac{1}{2}abc \gamma^{0}\psi.
\end{equation}
The primary constraints are given by
\begin{equation}  \label{Z}
p_{N}=\frac{\partial {\cal
L}}{\partial\dot{N}}=0,\hspace{.5cm}\pi_{\psi}=p_{\psi}-\frac{1}{2}abc\bar{\psi}\gamma^0=0,\hspace{.5cm}\pi_{\bar{\psi}}=p_{\bar{\psi}}+\frac{1}{2}
abc \gamma^0 \psi=0.
\end{equation}
In terms of the conjugate momenta the Hamiltonian is given by
\begin{equation}  \label{AB}
H=p_a \dot{a}+p_b \dot{b}+p_c \dot{c}+p_{\psi}\dot{\psi}+\dot{\bar{\psi}}p_{%
\bar{\psi}}+p_N \dot{N}-{\cal L},
\end{equation}
leading to
\begin{equation}  \label{AC}
H=-\frac{N}{4}\left(\frac{a}{bc}p_a^2 +\frac{b}{ac}p_b^2 +\frac{c}{ab}%
p_c^2\right)+\frac{N}{2}\left(\frac{p_a p_b}{c}+\frac{p_b p_c}{a}+\frac{p_a
p_c}{b}\right)-\Lambda N abc+NabcV(\bar{\psi},\psi).
\end{equation}
Because of the existence of constraints (\ref{Z}), the Lagrangian
of the system is singular and the total Hamiltonian can be
constructed by adding to $H$ the primary constraints multiplied by
arbitrary functions of time $\lambda(t)$, $\lambda_{\psi}(t)$ and
$\lambda_{\bar{\psi}}(t)$
\begin{eqnarray}  \label{AD}
H_T&=&-\frac{N}{4}\left(\frac{a}{bc}p_a^2 +\frac{b}{ac}p_b^2 +\frac{c}{ab}%
p_c^2\right)+\frac{N}{2}\left(\frac{p_a p_b}{c}+\frac{p_b
p_c}{a}+\frac{p_a p_c}{b}\right)\nonumber\\&-&\Lambda N
abc+NabcV(\bar{\psi},\psi)+\lambda
p_N+\pi_{\psi}\lambda_{\psi}+\lambda_{\bar{\psi}}\pi_{\bar{\psi}},
\end{eqnarray}
where the subscript $T$ stands for total. The requirement that the
primary constraints should hold during the evolution of the system
means that
\begin{equation}  \label{AE}
\dot{p_N}=\left\{p_N,H_T\right\}\approx 0,
\end{equation}and
\begin{equation}\label{rev 1}
\dot{\pi_{\psi}}=\left\{\pi_{\psi},H_T\right\}\approx
0,\hspace{.5cm}\dot{\pi_{\bar{\psi}}}=\left\{\pi_{\bar{\psi}},H_T\right\}\approx
0.\end{equation} The condition (\ref{AE}) leads to the secondary
constraint
\begin{equation}  \label{AF}
{\cal H}=-\frac{1}{4}\left(\frac{a}{bc}p_a^2 +\frac{b}{ac}p_b^2 +\frac{c}{ab}%
p_c^2\right)+\frac{1}{2}\left(\frac{p_a p_b}{c}+\frac{p_b
p_c}{a}+\frac{p_a p_c}{b}\right)-\Lambda
abc+abcV(\bar{\psi},\psi)=0,
\end{equation}while relations (\ref{rev 1}) only fix the
functions $\lambda_{\psi}$ and $\lambda_{\bar{\psi}}$
\begin{equation}\label{rev 2}
\lambda_{\psi}=-2N\gamma^0 \frac{\partial V}{\partial
\bar{\psi}},\hspace{.5cm}\lambda_{\bar{\psi}}=2N\frac{\partial
V}{\partial \psi}\gamma^0.
\end{equation}
The study of the algebra of  constraints (\ref{Z}) together with
the Poisson brackets of the secondary constraint with other
conjugate variables show that there are no additional constraints
in this dynamical system \cite{YU}.

Although the field equations obtained above can be solved after a
suitable form for the potential $V(\bar{\psi},\psi)$ has been
chosen \cite{24,25}, the Hamiltonian constraint (\ref{AF}) does
not have the desired form for the construction of the quantized
model or for the discussion of the problem of time in this model.
Thus, to transform Lagrangian (\ref{K}) to a more manageable form,
consider the following change of variables
\begin{equation}  \label{AG}
a=e^{u+v+\sqrt{3}w},\hspace{.5cm} b=e^{u+v-\sqrt{3}w},\hspace{.5cm}
c=e^{u-2v}.
\end{equation}
In terms of these new variables, Lagrangian (\ref{K}) takes the form
\begin{equation}  \label{AH}
{\cal L}=\frac{3}{N}\left(\dot{u}^2 -\dot{v}^2 -\dot{w}^2\right)e^{3u}+%
\Lambda N e^{3u}+\frac{1}{2}Ne^{3u}\left[\frac{1}{N}(\bar{\psi}\gamma^0 \dot{%
\psi}-\dot{\bar{\psi}}\gamma^0 \psi)-2V(\bar{\psi},\psi)\right].
\end{equation}
The momenta conjugate to $u$, $v$ and $w$ are
\begin{equation}  \label{AI}
p_u=\frac{\partial {\cal L}}{\partial \dot{u}}=\frac{6}{N}\dot{u}e^{3u},%
\hspace{.5cm} p_v=\frac{\partial {\cal L}}{\partial \dot{v}}=-\frac{6}{N}%
\dot{v}e^{3u},\hspace{.5cm} p_w=\frac{\partial {\cal L}}{\partial \dot{w}}=-%
\frac{6}{N}\dot{w}e^{3u},
\end{equation}
giving rise to the following Hamiltonian
\begin{equation}  \label{AJ}
H_T=\frac{N}{12}e^{-3u}\left(p_u^2 -p_v^2 -p_w^2 \right)+\left[V(\bar{\psi}%
,\psi)-\Lambda\right]Ne^{3u}+\lambda
p_N+\pi_{\psi}\lambda_{\psi}+\lambda_{\bar{\psi}}\pi_{\bar{\psi}},
\end{equation}
with the corresponding Hamiltonian constraint becoming
\begin{equation}  \label{AK}
{\cal H}=\frac{1}{12}e^{-3u}\left(p_u^2 -p_v^2 -p_w^2 \right)+\left[V(\bar{%
\psi},\psi)-\Lambda\right]e^{3u}=0.
\end{equation}
Now, variation of Lagrangian (\ref{AH}) with respect to its
dynamical variables yields the following field equations
\begin{equation}  \label{AL}
\dot{\psi}+\frac{3}{2}\dot{u}\psi+N\gamma^0 \frac{\partial V}{\partial \bar{%
\psi}}=0,
\end{equation}
\begin{equation}  \label{AM}
\dot{\bar{\psi}}+\frac{3}{2}\dot{u}\bar{\psi}-N\frac{\partial V}{\partial
\psi}\gamma^0=0,
\end{equation}
\begin{equation}  \label{AN}
\frac{2}{N}\ddot{u}+\frac{3}{N}\dot{u}^2- 2\frac{\dot{N}}{N^2}\dot{u}+\frac{3%
}{N}(\dot{v}^2+\dot{w}^2)-\Lambda N-\frac{1}{2}N\left[\frac{1}{N}(\bar{\psi}%
\gamma^0 \dot{\psi}-\dot{\bar{\psi}}\gamma^0 \psi)-2V(\bar{\psi},\psi)\right]%
=0,
\end{equation}
\begin{equation}  \label{AO}
\left(\frac{1}{N}\dot{v}e^{3u}\right)^{.}=0,
\end{equation}
\begin{equation}  \label{AP}
\left(\frac{1}{N}\dot{w}e^{3u}\right)^{.}=0,
\end{equation}
\begin{equation}  \label{AQ}
\frac{3}{N^2}\left(\dot{u}^2-\dot{v}^2-\dot{w}^2\right)+ \left(V(\bar{\psi}%
,\psi)-\Lambda\right)=0.
\end{equation}

Up to this point the cosmological model, in view of the concerning
issue of time, has been rather general and of course
under-determined. Before trying to solve these equations we must
decide on a choice of time in the theory. The general solutions of
the
system (\ref{AL})-(\ref{AQ}) for a free spinor field can be written as \cite%
{21}
\begin{equation}  \label{AR}
\bar{\psi}\psi=-\left[\frac{m}{-2\Lambda}+\sqrt{\frac{m^2}{4\Lambda^2}+ \frac{C^2}{%
-3\Lambda}}\cos\left(\sqrt{-3\Lambda} \int_{t_0}^{t}N(t^{\prime
})dt^{\prime }\right)\right]^{-1},
\end{equation}
\begin{equation}  \label{AS}
u(t)=\ln\left[\frac{m}{-2\Lambda}+\sqrt{\frac{m^2}{4\Lambda^2}+ \frac{C^2}{%
-3\Lambda}}\cos\left(\sqrt{-3\Lambda} \int_{t_0}^{t}N(t^{\prime })dt^{\prime
}\right)\right]^{1/3},
\end{equation}
\begin{equation}  \label{AT}
v(t)=\frac{c_1}{C}\ln\left|\frac{(B-A)\tan\left(\frac{1}{2}\sqrt{-3\Lambda}
\int_{t_0}^{t}N(t^{\prime })dt^{\prime }\right)+C/\sqrt{-3\Lambda}} {%
(B-A)\tan\left(\frac{1}{2}\sqrt{-3\Lambda} \int_{t_0}^{t}N(t^{\prime
})dt^{\prime }\right)-C/\sqrt{-3\Lambda}}\right|,
\end{equation}
\begin{equation}  \label{AU}
w(t)=\frac{c_2}{C}\ln\left|\frac{(B-A)\tan\left(\frac{1}{2} \sqrt{-3\Lambda}%
\int_{t_0}^{t}N(t^{\prime })dt^{\prime }\right)+C/\sqrt{-3\Lambda}} {%
(B-A)\tan\left(\frac{1}{2}\sqrt{-3\Lambda}\int_{t_0}^{t}N(t^{\prime
})dt^{\prime }\right) -C/\sqrt{-3\Lambda}}\right|,
\end{equation}
where $A=\frac{m}{-2\Lambda}$ and $B=\sqrt{\frac{m^2}{4\Lambda^2}+\frac{C^2}{%
-3\Lambda}}$ with $c_1$ and $c_2$ being two integrating constants such that $%
C^2=9(c_1^2+c_2^2)$. To obtain the above solutions we have assume that $%
\Lambda<0$. The corresponding solutions for $\Lambda>0$ can easily
be obtained by the replacement of the cosine  function in
(\ref{AS}) with its hyperbolic counterpart. The under-determinacy
problem at the classical level may be removed by using the gauge
freedom via fixing the gauge. For example, we can work in the
gauge $N=1$ which usually is chosen in classical cosmological
models and is called the cosmic time gauge. It is also worth
noting that with the form (\ref{A}) as the space-time metric, we
have done the first step in gauge fixing, namely $N^i=0$, where
$N^i$ is the shift vector. For other candidates in gauge fixing
see \cite{14} and \cite{16}. In any case, after fixing the gauge,
the lapse function $N(t)$ and shift vectors $N^i(t)$ are
eliminated from the field equations, rendering  them as the
Hamiltonian and momentum constraints. Elimination of the lapse
function from equations (\ref{AR})-(\ref{AU}) gives rise to the
intrinsic dynamics of the system as a relationship between the
3-geometry and matter field which is independent of the choice of
time in a particular gauge. The result is
\begin{equation}  \label{AV}
\bar{\psi}\psi=-e^{-3u},
\end{equation}%
or
\begin{equation}  \label{AW}
u(\bar{\psi}\psi)=\ln\left(\frac{-1}{\bar{\psi}\psi}\right)^{1/3},
\end{equation}
\begin{equation}  \label{AX}
v(\bar{\psi}\psi)=\frac{c_1}{C}\ln\left|\frac{\frac{B-A}{1+A\bar{\psi}\psi}%
\sqrt{[(B-A)\bar{\psi}\psi-1] [(B+A)\bar{\psi}\psi+1]} +C/\sqrt{-3\Lambda}}{%
\frac{B-A}{1+A\bar{\psi}\psi}\sqrt{[(B-A)\bar{\psi}\psi-1] [(B+A)\bar{\psi}%
\psi+1]} -C/\sqrt{-3\Lambda}}\right|,
\end{equation}
\begin{equation}  \label{AY}
w(\bar{\psi}\psi)=\frac{c_2}{C}\ln\left|\frac{\frac{B-A}{1+A\bar{\psi}\psi}%
\sqrt{[(B-A)\bar{\psi}\psi-1] [(B+A)\bar{\psi}\psi+1]} +C/\sqrt{-3\Lambda}}{%
\frac{B-A}{1+A\bar{\psi}\psi}\sqrt{[(B-A)\bar{\psi}\psi-1] [(B+A)\bar{\psi}%
\psi+1]} -C/\sqrt{-3\Lambda}}\right|.
\end{equation}

\section{Identification of time in the classical model}

In the definition of an intrinsic time the Hamiltonian constraint
plays a crucial role. In any constrained system we can impose the
constraints in different steps. In classical mechanics, for
example, we may first solve the equations of constraint to reduce
the degrees of freedom of the system and obtain a minimal number
of dynamical variables. On the other hand, we may multiply the
constraint by a variable parameter and add it to the Lagrangian.
This Lagrange multiplier plays the role of an additional dynamical
variable and the equations of motion consist of those obtained
from variation of the Lagrangian with respect to the dynamical
variables plus the equation of constraint. Solving this system of
equations of motion leads to either the time evolution of the
system or the forces of constraint. Also, when quantizing the
system, we may impose the constraint before or after the
quantization has been done. Now, if our system is the entire
universe, e.g. in the case of quantum cosmology, these procedures
result in different approaches to the problem of time
reparameterization. Here our goal is to reparameterize time in the
classical model in a manner described below. We first solve the
equation of constraint to obtain a set of genuine canonical
variables with which to construct the Hamiltonian. Equations of
motion are then obtained from this Hamiltonian and describe the
evolution of the system with respect to this intrinsic time.

To start, let us rewrite Lagrangian (\ref{AH}) in the following form
\begin{equation}  \label{AZ}
{\cal L}=\dot{u}p_u+\dot{v}p_v+\dot{w}p_w+p_{\psi}\dot{\psi}+ \dot{\bar{\psi}%
}p_{\bar{\psi}}+N\left[-\frac{1}{12}e^{-3u}(p_u^2-
p_v^2-p_w^2)+e^{3u}(\Lambda -V)\right].
\end{equation}
The role of the lapse function as a Lagrange multiplier is now
clear in the above form of the Lagrangian. The term in the square
brackets is indeed the Hamiltonian constraint. The procedure one
should follow is the same as that described in \cite{16}. First we
make a choice of time in terms of one of the dynamical variables
in the model and then solve the constraint equation (\ref{AK}) for
its conjugate momentum and substitute the result in the
Lagrangian. This process leads to a reduced Hamiltonian. A natural
choice for time in terms of the intrinsic geometry would be the
use of the scale factors. We may also take their conjugate momenta
or a scalar character of the matter field. These latter choices
are often referred to as extrinsic time.

Let us now examine the above procedure in the following cases. First,
suppose we choose $t=u$ as an intrinsic time. Solving the constraint
equation (\ref{AK}) for $p_u$ and substituting the result in (\ref{AZ})
yields the square-root Hamiltonian
\begin{equation}  \label{BA}
H=\left(p_v^2+p_w^2+12\Lambda e^{6t}-48mp_{\psi}p_{\bar{\psi}}\right)^{1/2}.
\end{equation}
Here, we have assumed that the spinor field is free with mass $m$,
{\it i.e.} $V(\bar{\psi},\psi)=m\bar{\psi}\psi$ and also used the
relation $\bar{ \psi}\psi=4e^{-6u}p_{\psi}p_{\bar{\psi}}$. Since
this Hamiltonian is independent of $v$, $w$, $\psi$ and
$\bar{\psi}$, these variables are cyclic
and their conjugate momenta are constant, that is $p_v=c_1$, $p_w=c_2$, $%
p_\psi=c_3$ and $p_{\bar{\psi}}=c_4$. The functional form of all
dynamical variables $v$, $w$, $\psi$ and $\bar{\psi}$ resulting
from (\ref{BA}) becomes
\begin{equation}  \label{BC}
\frac{1}{\sqrt{12|\Lambda|}} \ln\left|\frac{\sqrt{C\pm12|\Lambda| e^{6t}}+%
\sqrt{C}}{\sqrt{12|\Lambda|} e^{3t}}\right|,
\end{equation}
where $C^2=c_1^2+c_2^2-48mc_3c_4$, with the upper and lower signs
denoting positive and negative cosmological constants
respectively. A problem related to Hamiltonian (\ref{BA}) is that
it is a time-dependent function. Such Hamiltonians describe a
system which exchanges energy with the surrounding environment.
However, in the case of cosmology where the system under
consideration is the whole universe, a surrounding environment
does not have any meaningful interpretation. Therefore, such a
Hamiltonian and the corresponding time parameter do not seem to be
suitable unless the cosmological constant is zero. It can be
easily shown that in this case all dynamical variables have a
linear behavior with time. Another feasible alternative for
intrinsic time would be to take $t=v$. Following the same
procedure described above we find
\begin{equation}  \label{BD}
H=\left(p_u^2-p_w^2-12\Lambda e^{6u}+48mp_{\psi}p_{\bar{\psi}}\right)^{1/2},
\end{equation}
which does not suffer from being dependent on time. This Hamiltonian shows
that the variables $w$, $\psi$ and $\bar{\psi}$ are cyclic and their
conjugate momenta are constants, that is $p_w=c_1$, $p_{\psi}=c_2$ and $p_{%
\bar{\psi}}=c_3$. The corresponding cosmology is then obtained from the
solution of the equations of motion given by $\dot{u}=\frac{\partial H}{%
\partial p_u}$ etc. The result is
\begin{equation}  \label{BE}
u(t)=\ln\left(e^{-3u_0}-\frac{3\sqrt{12\Lambda}}{C}t\right)^{-1/3},
\end{equation}
\begin{equation}  \label{BF}
w(t)=\frac{c_1}{C}t+w_0,
\end{equation}
\begin{equation}  \label{BG}
\psi(t)=\frac{248mc_3}{C}t+\psi_0,
\end{equation}
where $u_0$, $w_0$ and $\psi_0$ are integrating constants and $%
C^2=48mc_2c_3-c_1^2$. The above expressions show that this choice
of the intrinsic time is appropriate for a universe with positive
or zero cosmological constant. In the case when $\Lambda=0$ all
canonical variables become cyclic and their time evolution becomes
linear with time, as in equations (\ref{BF}) and (\ref{BG}).
Choosing $w$ as time gives rise to the same results.

As we mentioned before, there are also time variables which are
conjugate to the intrinsic time known as the extrinsic time. For
example, consider the case where $t=p_u$. Solving the constraint
equation for $u$ leads us to the following Hamiltonian
\begin{equation}  \label{BH}
H=\frac{1}{\sqrt{12\Lambda}}\left(t^2-p_v^2-p_w^2+ 48mp_{\psi}p_{\bar{\psi}%
}\right)^{1/2}.
\end{equation}
This Hamiltonian is again time-dependent and describes a universe with
positive cosmological constant. Since the variables $v$, $w$, $\psi$ and $%
\bar{\psi}$ are all cyclic their corresponding momenta are constants, give
by $p_v=c_1$, $p_w=c_2$, $p_\psi=c_3$ and $p_{\bar{\psi}}=c_4$. It can be
shown that the evolution of all dynamical variables has the functional form
\begin{equation}  \label{BI}
\frac{1}{\sqrt{12\Lambda}}\ln\left(t+\sqrt{t^2+C^2}\right),
\end{equation}
where $C^2=48mc_3c_4-c_1^2-c_2^2$. Since the constraint equation
(\ref{AK}) is independent of $v$ and $w$ the choices $p_v$ and
$p_w$ do not make good as time parameters in our model. In
addition to the types of time described above, there is yet
another choice of the extrinsic time which corresponds to the
matter fields in the theory. In the case of a scalar field, the
natural choice for time is the scalar field $\phi$ itself, as has
been done in \cite{16}. In our model the matter field is a
4-spinor $\psi$ with complex components and the simplest scalar
constructed from it is $\bar{\psi}\psi$. Indeed, it is easy to see
that this quantity is proportional to the energy density of the
spinor field, ($\rho_{cl}=m\bar \psi_{cl}\psi_{cl}$), and
is the only observable that enters in the classical Einstein equations.
Thus, as the final time identification we take $t=\bar{%
\psi}\psi$ and are led to the following Hamiltonian
\begin{equation}  \label{BJ}
H=\frac{1}{\sqrt{48m}}\left(12\Lambda e^{6u}-p_u^2+p_v^2+p_w^2\right)^{1/2}.
\end{equation}
The classical cosmology resulting from this Hamiltonian can be
read from the equations of motion $\dot{u}=\frac{\partial
H}{\partial p_u}$ etc., with the result
\begin{equation}  \label{BK}
u(t)=\ln\left(\frac{3}{C}\sqrt{\frac{\Lambda}{4m}}t+ e^{-3u_0}\right)^{-1/3},
\end{equation}
\begin{equation}  \label{BL}
v(t)=\frac{c_1}{C\sqrt{48m}}t+v_0,\hspace{.5cm} w(t)=\frac{c_2}{C\sqrt{48m}}%
t+w_0,
\end{equation}
\begin{equation}  \label{BM}
p_u(t)=\sqrt{12\Lambda}\left(\frac{3}{C}\sqrt{\frac{\Lambda}{4m}}t+
e^{-3u_0}\right)^{-1},\hspace{.5cm} p_v=c_1,\hspace{.5cm}p_w=c_2,
\end{equation}
where $c_1$, $c_2$, $v_0$, $w_0$ and $u_0$ are integrating constants and $%
C^2=c_1^2+c_2^2$. It is clear from the above equations that this type of
time is suitable for a universe with positive or zero cosmological constant.
Again in the case of a zero cosmological constant the dynamical variables
have a linear behavior with time.

Our classical investigation on the problem of time is now complete. In the
next section we shall pursue this problem within the context of a quantum
cosmological model.

\section{Quantization of the model}

The usual approach to canonical quantization of a cosmological
model is the Wheeler-DeWitt approach where one uses the Dirac
method to quantize the degrees of freedom of the system. The role
of constraints in their operator form is to annihilate the wave
function of the universe. This procedure leads one to the basic
equation of quantum cosmology, the so called WD equation. This
approach to quantum cosmology has its own problems and is not the
subject of study in this paper, see \cite{12} and \cite{15}.
However, as was done in the previous section, one may solve the
constraint before using it in the theory, in particular before
quantizing the system. If we do so, we are led to the
Schr\"{o}dinger equation
\begin{equation}  \label{BN}
H\Psi=i\frac{\partial \Psi}{\partial t},
\end{equation}
where $H$ is the operator form of the reduced Hamiltonian, {\it
i.e.} the operator form of one of the Hamiltonians given in
(\ref{BA}), (\ref{BD}), (\ref{BH}) and (\ref{BJ}). There are,
however, some problems related to these Hamiltonians at the
quantum level. Firstly, they are all given as square-roots and
thus by means of the spectral theorem are assumed to be positive
definite and Hermitian operators. Secondly, the Schr\"{o}dinger
equation (\ref{BN}) is in general a time-dependent equation
because of the dependence of $H$ on $t$. Thus one should not
conclude that this equation is equivalent to a WD or a second
order Klein-Gordon type equation like
\begin{equation}  \label{BO}
H^2\Psi=-\frac{\partial^2 \Psi}{\partial t^2},
\end{equation}
to remove the square-root form of the Hamiltonian. Indeed, by acting with $H$
on both sides of equation (\ref{BN}) one obtains
\begin{equation}  \label{BP}
H^2\Psi=-\frac{\partial^2 \Psi}{\partial t^2}-i\frac{\partial H}{\partial t}%
\Psi,
\end{equation}
which, of course, has different solutions from those given by equation (\ref%
{BO}), unless the Hamiltonian is time-independent. For a careful
description of the details of this issue see \cite{14,16}.
However, as is well known from the elementary quantum mechanics,
if the Hamiltonian is time-independent, the solutions of the
Schr\"{o}dinger equation can be written as
\begin{equation}  \label{BQ}
\Psi(\vec{x},t)=\exp\left[-iH(t-t_0)\right]\Psi(\vec{x},t_0).
\end{equation}
In the case of a time-dependent Hamiltonian the above formula
should be modified by the Dyson series
\begin{equation}  \label{BR}
\Psi(\vec{x},t)=\left[1+\sum_{n=1}^{\infty}(-i)^{n}\int_{t_0}^tdt_{1}
\int_{t_0}^{t_1}dt_{2}\cdots\int_{t_0}^{t_{n-1}}dt_{n}H(t_1) H(t_2)\cdots
H(t_n)\right]\Psi(\vec{x},t_0).
\end{equation}
In practice, interesting cases happen when
\begin{equation}  \label{BS}
\left[H(t^{\prime }),H(t^{\prime \prime })\right]=0,
\end{equation}
in which case the Dyson series becomes
\begin{equation}  \label{BI}
\Psi(\vec{x},t)=\left[\exp\left(-i\int_{t_0}^{t}H(t^{\prime })dt^{\prime
}\right)\right]\Psi(\vec{x},t_0).
\end{equation}
Another feature of relation (\ref{BS}) is that the Hamiltonian has
the same eigenstates at all times, in other words if
$\Psi_{E_0}(\vec{x})$ is
the eigenstate of $H(t_0)$ at some initial time $t_0$ : $H(t_0)\Psi_{E_0}(%
\vec{x})=E_0\Psi_{E_0}(\vec{x})$, then $\Psi_{E_0}(\vec{x})$ is
also the eigenstate of $H(t)$ with another eigenvalue $E(t)$ at
time $t$
\begin{equation}  \label{BU}
H(t)\Psi_{E_0}(\vec{x})=E(t)\Psi_{E_0}(\vec{x}).
\end{equation}
In conclusion we are led to the following relationship for the time
evolution of the wave function
\begin{equation}  \label{BV}
\Psi(\vec{x},t)=\left[\exp\left(-i\int_{t_0}^{t}E(t^{\prime })dt^{\prime
}\right)\right]\Psi_{E_0}(\vec{x},t_0).
\end{equation}
To find $\Psi_{E_0}(\vec{x},t_0)$ we can use the eigenvalue equation $%
H(t_0)\Psi_{E_0}(\vec{x})=E_0\Psi_{E_0}(\vec{x})$ in the form
\begin{equation}  \label{BW}
H^2(t_0)\Psi_{E_0}(\vec{x})=E_0^2\Psi_{E_0}(\vec{x}).
\end{equation}
This is an allowed equation since $H$ is evaluated at a particular fixed
time $t_0$ and thus the square-root problem is now resolved. In view of
dealing with the Klein-Gordon or WD equation in our Schr\"{o}%
dinger approach, we do not need to be concerned with the inner
product of states since the determination of the wave functions in
the Schr\"{o}dinger equation and the construction of the Hilbert
space proceed in the usual sense of quantum mechanics. The final
remark about the reduced Hamiltonian in the above procedure is the
factor-ordering problem when one embarks on constructing a quantum
mechanical operator equation. In the class of Hamiltonians
represented by (\ref{BA}) and (\ref{BH}), this problem is not too
important since there is no presence of a canonical variable with
its conjugate momentum. However, consider for example, Hamiltonian
(\ref{BD}) in which both $u$ and its conjugate momentum $p_u$ are
present and do not commute. In dealing with such Hamiltonians at
the quantum level care should be taken when one tries to make a
change of variable. This is so because such an operation may
change the form of the Hamiltonian where a product of a variable
with its conjugate momentum has now appeared. This is an
indication that in quantizing the system, the ordering problem
becomes important. As an example, consider a change of variable
$r^2=e^{3u}$ applied to Hamiltonian (\ref{BD}), changing it to
\begin{equation}  \label{BX}
H=\left(\frac{9}{4}r^2p_r^2-p_w^2+48mp_{\psi}p_{\bar{\psi}}-12\Lambda
r^4\right)^{1/2}.
\end{equation}
Under the same change of variable Hamiltonian (\ref{BJ}) becomes
\begin{equation}  \label{BY}
H=\left[-\frac{3}{64m}r^2p_r^2+\frac{1}{48m}\left(p_v^2+p_w^2\right)+ \frac{%
\Lambda}{4m}r^4\right]^{1/2}.
\end{equation}
It is now clear that in replacing $r$ and $p_r$ with their
corresponding operators, the ordering considerations should be
taken into account. We return to this issue again in the next
section.

\section{Time identification in the quantized model}

Let us now use the above theory in the problem at hand and investigate time
reparameterization introduced in the last section in the context of the
quantized model. In the case when time is chosen as $t=u$, the Hamiltonian
is given by (\ref{BA}) satisfying the relation (\ref{BS}) for all times. To
find the wave function $\Psi(v,w,\bar{\psi}\psi,t)$, we must first solve
equation (\ref{BW}) for this Hamiltonian. With the usual replacement $%
p_v\rightarrow -i\frac{\partial}{\partial v}$ and similarly for $p_w$, $%
p_{\psi}$ and $p_{\bar{\psi}}$ this equation becomes
\begin{equation}  \label{BZ}
\left[-\frac{\partial^2}{\partial v^2}-\frac{\partial^2}{\partial w^2}%
+12\Lambda e^{6t_0}+48m\frac{\partial^2}{\partial \psi \partial \bar{\psi}}%
\right]\Psi_{E_0}(v,w,\bar{\psi}\psi)=E_0^2\Psi_{E_0}(v,w,\bar{\psi}\psi).
\end{equation}
The solutions of the above differential equation are separable and may be
written in the form $\Psi(v,w,\bar{\psi}\psi)=V(v)W(w)f(\bar{\psi}\psi)$
leading to
\begin{equation}  \label{CA}
\frac{1}{V}\frac{d^2V}{dv^2}=\alpha^2,\hspace{.5cm}\frac{1}{W}\frac{d^2W}{%
dw^2}=\beta^2, \\
\end{equation}
\begin{equation}  \label{CB}
\frac{1}{f}\frac{\partial^2 f}{\partial \psi \partial \bar{\psi}}=\gamma^2,
\end{equation}
where $\alpha$, $\beta$ and $\gamma$ are separation constants satisfying $%
\alpha^2+\beta^2=48m \gamma^2+12\Lambda e^{6t_0}-E_0^2$. Equations (\ref{CA}%
) have simple solutions in the form of exponential functions
$e^{-|\alpha|v}$ and $e^{-|\beta|w}$. The exponents are chosen so
that $\Psi(v,w\rightarrow + \infty)=0$. To find the solutions of
equation (\ref{CB}) we use the ansatz
\begin{equation}  \label{AA}
f(\bar{\psi}\psi)=\sum_{n=0}^{\infty}c_n(\bar{\psi}\psi)^n.
\end{equation}
After a little algebra we find $c_n=\frac{\gamma^{2n}}{(n!)^2}c_0$ and
\begin{equation}  \label{BB}
f(\bar{\psi}\psi)=\sum_{n=0}^{\infty}\frac{\gamma^{2n}}{(n!)^2}(\bar{\psi}%
\psi)^n.
\end{equation}
It is easy to check that this series converges for all value of $\bar{\psi}%
\psi$. We may thus write the solution of equation (\ref{BZ}) as
\begin{equation}  \label{CD}
\Psi_{E_0}(v,w,\bar{\psi}\psi)=e^{-|\alpha|v-|\beta|w} \sum_{n=0}^{\infty}%
\frac{\gamma^{2n}}{(n!)^2}(\bar{\psi}\psi)^n.
\end{equation}
According to equation (\ref{BU}) the above eigenfunctions should
also be the eigenfunctions of $H(t)$ given by (\ref{BA}) with
eigenvalues $E(t)$ such that
\begin{equation}  \label{CE}
E(t)=\left[12\Lambda\left(e^{6t}-e^{6t_0}\right)+E_0^2\right]^{1/2}.
\end{equation}
The time evolution of the wave function is then given by
(\ref{BV}) with the result
\begin{eqnarray}
\Psi(v,w,\bar{\psi}\psi,t)&=&\left[\exp\left(-i\int_{t_0}^{t}E(t^{\prime
})dt^{\prime }\right)\right] \Psi_{E_0}(v,w,\bar{\psi}\psi)  \nonumber \\
&=&\left\{\exp\left(-i\int_{t_0}^{t}\left[12\Lambda(e^{6t^{\prime
}}-e^{6t_0})+E_0^2\right]^{1/2}dt^{\prime }\right)\right\} \Psi_{E_0}(v,w,%
\bar{\psi}\psi)  \nonumber \\
&=&\left\{\exp\frac{1}{72\Lambda}\left[ 2\left[12\Lambda(e^{6t}-e^{6t_0})+E_0^2\right]%
^{1/2}-2E_0+(E_0^2-12\Lambda e^{6t_0}) \times \right.\right. \nonumber   \\
& &\left. \left. \ln\left|\frac{\left\{\left[12\Lambda(e^{6t}-e^{6t_0})+E_0^2\right]%
^{1/2}-(E_0^2-12\Lambda e^{6t_0})^{1/2}\right\}\left\{E_0+(E_0^2-12\Lambda
e^{6t_0})^{1/2}\right\}}{\left\{\left[12\Lambda(e^{6t}-e^{6t_0})+E_0^2\right]%
^{1/2}+(E_0^2-12\Lambda e^{6t_0})^{1/2}\right\}\left\{E_0-(E_0^2-12\Lambda
e^{6t_0})^{1/2}\right\}}\right| \right] \right\} \times  \nonumber \\
& &\Psi_{E_0}(v,w,\bar{\psi}\psi).  \label{CF}
\end{eqnarray}
A comment about the Hamiltonian (\ref{BA}) is that its
eigenenergies have a continuous spectrum in the range
$[\sqrt{12\Lambda}e^{3t},+\infty)$ which for, $\Lambda\geq0$, is
positive definite as required. This means that our choice of the
time parameter is suitable for a universe with positive or zero
cosmological constant. However, in the case of $\Lambda=0$, the
time evolution of the wave function is simply given by (\ref{BQ}).
Note that in the classical model in this gauge both positive and
negative cosmological constants were valid.

Let us now deal with the choice $t=v$ as time. In this case the
Hamiltonian is given by (\ref{BD}). However, to deal with the
ordering problem we work with Hamiltonian (\ref{BX}). To guarantee
Hermiticity, the operator form corresponding to this Hamiltonian
should be written as
\begin{equation}  \label{CG}
H=\left(\frac{9}{4}r^{p}p_{r}r^{2-2p}p_{r}r^{p}-p_w^2+48mp_{\psi}p_{\bar{\psi%
}}-12\Lambda r^4\right)^{1/2},
\end{equation}
where the parameter $p$ denotes the ambiguity in the ordering of
factors $r$ and $p_r$ in the first term of (\ref{BX}). Taking
$p=0$, equation (\ref{BW}) reads
\begin{equation}  \label{CH}
\left(-\frac{9}{4}\frac{\partial}{\partial r}r^2\frac{\partial}{\partial r}+%
\frac{\partial^2}{\partial w^2}-48m\frac{\partial^2}{\partial \psi \partial
\bar{\psi}}-12\Lambda r^4\right)\Psi_{E_0}(r,w,\bar{\psi}\psi)=E_0^2%
\Psi_{E_0}(r,w,\bar{\psi}\psi).
\end{equation}
We again write the solutions in the form $\Psi(r,w,\bar{\psi}%
\psi)=R(r)W(w)f(\bar{\psi}\psi)$ and find
\begin{equation}  \label{CI}
\frac{1}{W}\frac{d^2W}{dw^2}=\alpha^2,
\end{equation}
\begin{equation}  \label{CJ}
\frac{1}{f}\frac{\partial^2 f}{\partial \psi \partial \bar{\psi}}=\frac{1}{%
48m}\beta^2,
\end{equation}
\begin{equation}  \label{CK}
\frac{d^2R}{dr^2}+\frac{2}{r}\frac{dR}{dr}+\frac{4}{9}\left(\frac{%
E_0^2-\gamma^2}{r^2}+12\Lambda r^2\right)R=0,
\end{equation}
where $\alpha$ and $\beta$ are separating constants and $\gamma^2=\alpha^2-%
\beta^2$. The solutions of equations (\ref{CI}) and (\ref{CJ}) are known as $$%
W(w)=e^{-|\alpha |w}\hspace{5mm} \mbox{and}\hspace{3mm} f(\bar{\psi}\psi)=\sum_{n=0}^{\infty}\frac{(\beta/%
\sqrt{48m})^{2n}}{(n!)^2}(\bar{\psi}\psi)^n.$$
Also, the solutions of (\ref%
{CK}) for a positive cosmological constant can be written in terms of Bessel
functions as
\begin{equation}  \label{CL}
R(r)=r^{-1/2} J_{\pm \nu}\left(\frac{2}{3}\sqrt{3\Lambda}r^2\right),
\end{equation}
where $\nu^2=1/16-(E_0^2-\gamma^2)/9$. The complete solution of equation (%
\ref{CH}) now reads
\begin{equation}  \label{CN}
\Psi_{E_0}(r,w,\bar{\psi}\psi)=e^{-|\alpha|w} r^{-1/2}J_{\nu}\left(\frac{2}{3%
}\sqrt{3\Lambda}r^2\right) \sum_{n=0}^{\infty}\frac{\left(\beta/\sqrt{48m}\right)^{2n}}{%
(n!)^2}(\bar{\psi}\psi)^n,
\end{equation}
where to avoid diverging solutions at $r=0$ we have removed the function $%
J_{-\nu}$. As is clear from the above equation the
wave function satisfies $\Psi(w\rightarrow + \infty)=0$ and $%
\Psi(r\rightarrow +\infty)=0$. In the limit $r\rightarrow 0$ the
wave function $\Psi(r,w,\bar{\psi}\psi)$ behaves as $r^{2\nu-1/2}$
and thus to have regular solutions near $r=0$ we must have
$2\nu-1/2\geq 0$. This condition restricts the eigenenergies to
the interval $[0,\gamma]$ and  the initial condition to
$\Psi(r=0)=0$ for the wave function as suggested in \cite{1}.
Contrary to the classical case where such a choice for the time
parameter was appropriate only for a positive cosmological
constant, $\Lambda$ can be negative as well as positive here. In
the case of a negative cosmological constant, the Bessel function
in equation (\ref{CL}) must be replaced with the modified Bessel
functions $K_\nu(x)$ and $I_\nu(x)$.
To satisfy $\Psi(r\rightarrow \infty)=0$ we restrict ourselves to functions $%
K_\nu(x)$. Thus for a negative cosmological constant
\begin{equation}  \label{ZZ}
\Psi_{E_0}(r,w,\bar{\psi}\psi)=e^{-|\alpha|w}r^{-1/2}K_{\nu}\left(\frac{2}{3}
\sqrt{3|\Lambda|}r^2\right)\sum_{n=0}^{\infty}\frac{\left(\beta/\sqrt{48m}\right)^{2n}}{%
(n!)^2} (\bar{\psi}\psi)^n.
\end{equation}
Again to avoid singularity at $r=0$ the order of the function
$K_\nu(x)$ should be pure imaginary; $\nu^2<0$ \cite{26}, which
results in the interval $(\sqrt{\gamma^2+9/16},+\infty)$ for the
allowed eigenenergies. One should note that it is impossible for
the wave function to satisfy the condition $\Psi(r=0)=0$ in this
case.

For a universe with zero cosmological constant the solutions of
equation (\ref{CK}) can be written as
\begin{equation}  \label{ZA}
R(r)\sim r^{c_+},r^{c_-},
\end{equation}
where $$c_{\pm}=\frac{1}{2}\left(-1\pm
\sqrt{1-16(E_0^2-\gamma^2)/9}\right).$$ Thus for the energies in
the interval $0\leq E_0\leq \gamma$ we can have
regular solutions $r^{c_+}$ satisfying $\Psi(r=0)=0$, while for $\sqrt{%
\gamma^2+9/16}\leq E_0< +\infty$ we are led to  oscillatory
solutions. Since Hamiltonian (\ref{CG}) is time-independent the
time evolution of $\Psi(\vec{x})$ is given by (\ref{BQ}) with the
result
\begin{equation}  \label{CO}
\Psi(r,w,\bar{\psi}\psi,t)=e^{-iE_0(t-t_0)}\Psi_{E_0}(r,w,\bar{\psi}\psi).
\end{equation}
The other choice for time studied in the last section is $t=p_u$
which may be interpreted as the conjugate to the choice $t=u$. It
can easily be shown that in this case the eigenstates of
Hamiltonian (\ref{BH}) are obtained by
interchanging the roles of $t_0$ and $E_0$ in equation (\ref{CF}) \cite%
{16}.

Our final discussion about the problem of time is to take a scalar
character of matter as the time parameter. As is well known the
simplest scalar which can be constructed out of a spinor field is
$\bar{\psi}\psi$. This suggests the choice $t=\bar{\psi}\psi$. In
this case we must solve equation (\ref{BW}) with Hamiltonian
(\ref{BJ}) or (\ref{BY}) to deal with the factor ordering.
With the same factor ordering parameter as we have taken in equation (\ref%
{CG}) we write
\begin{equation}  \label{CP}
\left[\frac{\partial}{\partial r}r^2\frac{\partial}{\partial r}-\frac{4}{9}%
\left(\frac{\partial^2}{\partial v^2}+\frac{\partial^2}{\partial w^2}\right)+%
\frac{16\Lambda}{3}r^4\right]\Psi_{E_0}(r,v,w)=\frac{64}{3}%
mE_0^2\Psi_{E_0}(r,v,w),
\end{equation}
with solution
\begin{equation}  \label{CX}
\Psi_{E_0}(r,v,w)=e^{-3/2(|\alpha|v+|\beta|w)}r^{-1/2}J_{\nu}\left(\frac{2}{3}%
\sqrt{3\Lambda}r^2\right),
\end{equation}
where $\nu^2=1/16+(64mE_0^2+3\gamma^2)/12$ and
$\gamma^2=\alpha^2+\beta^2$. The same analysis done for equation
(\ref{CN}) suggests that if $\Lambda\geq 0$ for all energies in
the interval $[0,+\infty)$, we have solutions satisfying the
condition $\Psi(r=0)=0$. In the case of a negative cosmological we
get solutions that are neither regular as $r\rightarrow 0$ nor in
the limit $r\rightarrow \infty$. Thus, as in the classical case
this choice of time is only suitable for a universe with
$\Lambda\geq 0$.

\section{Conclusions}

In this paper we have studied the time reparameterization problem
in a Bianchi type I cosmology with a Dirac spinor field as the
matter source at both the classical and quantum levels. This
problem arises from the fact that Einstein field equations in
classical general relativity are under-determined, resulting in
the requirement of imposing a gauge condition before a solution
can be found. In the ADM formalism of general relativity these
gauge conditions are those that are commonly imposed on the lapse
function or shift vectors.

With line element (\ref{A}) we have taken the shift vectors
$N^i=0$ so that the only gauge freedom in our model relates to
different choices of lapse function $N(t)$, giving rise to
different choices of time parameters in the model. At the quantum
level however, the gauge freedom appears in the form of the
Hamiltonian and momentum constraints. Again, because of the form
of our metric the momentum constraint is automatically satisfied
and one should only deal with the Hamiltonian constraint. In order
to fix the gauge, we chose a time parameter defined in terms of
the 3-geometry (intrinsic time) or its conjugate momentum. We also
explored the possibility of the matter field playing such a role
(extrinsic time). The Hamiltonian constraint for the conjugate
variables was then solved and the solutions, considered as time,
were used to find a reduced Hamiltonian and consequently the
classical field equations. Of course, these equations do not
suffer from the under-determinacy problem. In this sense we have
taken three intrinsic times as $t=u,v,w$ where $u$, $v$ and $w$
are related to the scale factors through equations (\ref{AG}). The
corresponding cosmologies in terms of these time parameters are
given by the relations (\ref{BC}) and (\ref{BE})-(\ref{BG}). These
solutions are valid for an arbitrary cosmological constant when
$t=u$ and a positive or zero cosmological constant in the case of
$t=v$ or $w$ respectively. Also, the extrinsic times, $t=p_u$ and
$t=\bar{\psi}\psi$, give rise to the classical cosmologies
(\ref{BI}) and (\ref{BK})-(\ref{BM}),  describing a universe with
a positive, negative or zero cosmological constant respectively.
We have also shown that $p_v$ and $p_w$ are not suitable for
playing the role of time. To quantize the model we have followed
the procedure introduced in \cite{16}. After fixing the time we
used the operator form of the reduced Hamiltonian to quantize the
system in the Schr\"{o}dinger picture. This procedure has led us
to a quantum cosmology with time dependent wave function
(\ref{CF}) in the
case $t=u$ which is appropriate for $\Lambda\geq 0$, or (\ref{CN})-(\ref{CO}%
) when $t=v$ which is suitable for a cosmological constant with
arbitrary sign. Finally, in the case where $\bar{\psi}\psi$ was
considered as the time variable we have obtained the wave function
(\ref{CX}) for a positive or zero cosmological
constant.\vspace{5mm}\newline \noindent {\bf
Acknowledgement}\vspace{2mm}\noindent\newline The authors would
like to thank the research council of Shahid Beheshti University
for financial support.


\begin{thebibliography}{99}
\bibitem{1} B. S. DeWitt, {\it Phys. Rev.} {\bf 160} (1967) 1113
\bibitem{2} C. W. Misner, {\it Phys. Rev.} {\bf 186} (1969) 1319
\bibitem{3} A. Vilenkin, {\it Phys. Lett.} B {\bf 117} (1982) 25
\bibitem{4} A. Vilenkin, {\it Phys. Rev.} D {\bf 27} (1983) 2848
\bibitem{5} A. Vilenkin, {\it Phys. Rev.} D {\bf 33} (1986) 3560
\bibitem{6} A. Vilenkin, {\it Phys. Rev.} D {\bf 37} (1988) 888
\bibitem{7} A. Vilenkin, {\it Phys. Rev.} D {\bf 50} (1994) 2581
\bibitem{8} J. B. Hartle and S. W. Hawking, {\it Phys. Rev.} D {\bf 28}
(1983) 2960
\bibitem{9} S. W. Hawking, {\it Nucl. Phys.} B {\bf 239} (1984) 257
\bibitem{10} J. J. Halliwell and S. W. Hawking, {\it Phys. Rev.} D {\bf 31}
(1985) 1777
\bibitem{11} G. W. Gibbons and J. B. Hartle, {\it Phys. Rev.} D {\bf 42}
(1990) 2458
\bibitem{12} D. L. Wiltshire, {\it An Introduction to Quantum Cosmology},
(gr-qc/0101003)
\bibitem{13} M. P. Ryan, {\it Hamiltonian Cosmology} Springer, Berlin (1972)%
\newline
M. P. Ryan and L. C. Shepley, {\it Homogeneous Relativistic Cosmologies},
Princeton University Press, Princeton (1975)
\bibitem{14} C. J. Isham, {\it Canonical quantum gravity and the problem of
time}, lectures presented at NATO Advanced Study Institute, Salamanca,
(1992) (gr-qc/9210011)
\bibitem{e} Proceedings of Nato Advanced Research Workshop:
"Physical Origins of Time asymmetry", edited by J. J. Halliwell,
Cambridge University Press, (1994)
\bibitem{15} T. P. Shestakova and C. Simeone, {\it Grav. Cosmol.} {\bf 10}
(2004) 161, (gr-qc/0409114) \newline
T. P. Shestakova and C. Simeone, {\it Grav. Cosmol.} {\bf 10} (2004) 257,
(gr-qc/0409119)
\bibitem{16} W. F. Blyth and C. J. Isham, {\it Phys. Rev.} D {\bf 11} (1975)
768
\bibitem{17} D. C. Salisbury, J. Helpert and A. Schmitz, (gr-qc/0503014)
\bibitem{18} J. M. Pons and D. C. Salisbury, {\it Phys. Rev.} D {\bf 71}
(2005) 124012, (gr-qc/0503013)
\bibitem{19} C. J. Isham and J. E. Nelson, {\it Phys. Rev.} D {\bf 10}
(1974) 3226
\bibitem{a} J. E. Nelson and C. Teitelboim, {\it Phys. Lett.} B {\bf
69} (1977) 81
\bibitem{b} M. Henneaux, {\it Gen. Rel. Grav.} {\bf 9} (1978) 1031
\newline
M. Henneaux, {\it Phys. Rev.} D {\bf 21} (1980) 857
\bibitem{c} P. D. D'Eath and J. J Halliwell, {\it Phys. Rev.} D {\bf
35} (1987) 1100
\bibitem{20} B. Vakili, S. Jalalzadeh and H. R. Sepangi, {\it J. Cosmol.
Astropart. Phys.} JCAP 05 (2005) 006, (gr-qc/0502076)
\bibitem{21} B. Vakili and H. R. Sepangi, {\it J. Cosmol. Astropart. Phys.}
JCAP 09 (2005) 008, (gr-qc/0508090)
\bibitem{22} C. Armendariz-Picon and P. B. Greene, {\it Gen. Rel. Grav.}
{\bf 35} (2003) 1637, (hep-th/0301129)
\bibitem{d} S. Hervik, {\it Class. Quantum Grav.} {\bf 17} (2000)
2765, (gr-qc/0003084)\newline T. Christodoulakis, T. Gakis and G. O.
Papadopoulos, {\it Class. Quantum Grav.} {\bf 19} (2002) 1013,
(gr-qc/0106065)
\bibitem{23} J. M. Jauch and F. Rohrlich, {\it Theory of Photons and
Electrons}, Springer, New York, 1976
\bibitem{YU} A. M. Khvedelidze and Yu. G. Palii, {\it Class.
Quantum Grav.} {\bf 18} (2001) 1767, (gr-qc/0103048)
\bibitem{24} B. Saha, {\it Phys. Rev.} D {\bf 64} (2001) 123501,
(gr-qc/0107013)
\bibitem{25} B. Saha, {\it Phys. Rev.} D {\bf 69} (2004) 124006,
(gr-qc/0308088)
\bibitem{26} W. Magnus, F. Oberhettinger and R. P. Soni, {\it Formulas and
Theorems for the Special Functions of Mathematical Physics},
Springer-Verlag, Berlin, 1966
\end{thebibliography}
\end{document}